\documentclass[conference]{IEEEtran}
\IEEEoverridecommandlockouts
\usepackage{cite}
\usepackage{amsmath,amssymb,amsfonts}
\usepackage{algorithm}
\usepackage{algorithmic}
\usepackage{graphicx}
\usepackage{textcomp}
\usepackage{float}
\usepackage{xcolor}
\usepackage{comment}
\usepackage[letterpaper, left=0.75in, right=0.75in, top=0.75in, bottom=1.2in]{geometry}
\usepackage{amsmath,amssymb,amsfonts}
\usepackage{algorithmic}
\usepackage{lipsum,graphicx}
\usepackage{textcomp}
\usepackage{xcolor}
\usepackage{caption}
\usepackage{tabularx}
\usepackage{subcaption}
\usepackage{acro}[=v2]
\usepackage{url}
\usepackage{algorithm}

\usepackage{makecell}
\usepackage{array}
\usepackage{textcomp}
\usepackage{stfloats}
\usepackage{url}
\usepackage{verbatim}

\usepackage{color,soul}
\usepackage{lipsum}
\usepackage{multirow}
\usepackage{tabularx,booktabs}

\usepackage{csquotes}

\usepackage{amsmath}
\usepackage{algorithm}
\usepackage{algorithmic}

\usepackage{array}
\usepackage{booktabs}
\usepackage{caption}
\usepackage{longtable}

\usepackage[caption=false,font=small]{subfig}
\usepackage{textcomp}
\usepackage{stfloats}
\usepackage{url}
\usepackage{verbatim}
\usepackage{graphicx}
\usepackage{algorithm}
\usepackage{algorithmic}

\usepackage{cite}
\usepackage{amsmath,amssymb,amsfonts}

\usepackage{csquotes}
\usepackage{xcolor}
\usepackage{graphicx}

\usepackage{color,soul}
\usepackage{lipsum}
\usepackage{multirow}
\usepackage{tabularx,booktabs}
\let\svq"
\catcode`"=\active
\def"{\texttt{\svq}}

\def\BibTeX{{\rm B\kern-.05em{\sc i\kern-.025em b}\kern-.08em
    T\kern-.1667em\lower.7ex\hbox{E}\kern-.125emX}}
    
\begin{document}

\title{SAJD: Self-Adaptive Jamming Attack Detection in AI/ML Integrated 5G O-RAN Networks}
\author{%
Md Habibur Rahman\textsuperscript{1,*},
Md Sharif Hossen\textsuperscript{2,*},
Nathan H. Stephenson\textsuperscript{3},
Vijay K. Shah\textsuperscript{2},
Aloizio Da Silva\textsuperscript{1}\\[0.25em]
\textit{\textsuperscript{1} Commonwealth Cyber Initiative, Virginia Tech, VA, USA, Emails: \{mhrahman, aloiziops\}@vt.edu}\\
\textit{\textsuperscript{2} NextG Wireless Lab, North Carolina State University, NC, USA, Emails: \{mhossen, vijay.shah\}@ncsu.edu}\\
\textit{\textsuperscript{3} George Mason University, VA, USA, Email: nstephe7@gmu.edu}%
\vspace{-0.19in}
\thanks{\textsuperscript{*}Both Authors contributed equally.}
}
%\vspace{-0.2in}
\maketitle

%\thanks{*Both authors have contributed equally.}
\begin{abstract}
The open radio access network (O-RAN) enables modular, intelligent, and programmable 5G network architectures through the adoption of software-defined networking (SDN), network function virtualization (NFV), and implementation of standardized open interfaces. It also facilitates closed-loop control and (non/near) real-time optimization of radio access network (RAN) through the integration of non-real-time applications (rApps) and near-real-time applications (xApps). However, one of the security concerns for O-RAN that can severely undermine network performance and subject it to a prominent threat to the security \& reliability of O-RAN networks is \textit{jamming attacks}. To address this, we introduce SAJD -- a self-adaptive jammer detection framework that autonomously detects jamming attacks in artificial intelligence (AI) / machine learning (ML)-integrated O-RAN environments. The SAJD framework forms a closed-loop system that includes near-real-time inference of radio signal jamming interference via our developed ML-based xApp, as well as continuous monitoring and retraining pipelines through rApps. Specifically, a labeler rApp is developed that uses live telemetry (i.e., KPIs) to detect model drift, triggers unsupervised data labeling, executes model training/retraining using the integrated \& open-source ClearML framework, and updates deployed models on the fly, without service disruption. Experiments on O-RAN-compliant testbed demonstrate that the SAJD framework outperforms state-of-the-art (offline-trained with manual labels) jamming detection approach in accuracy and adaptability under various dynamic and previously unseen interference scenarios.

%\vspace{-0.5mm}
\begin{comment}
The Open Radio Access Network (O-RAN) enables modular, intelligent, and programmable 5G radio access network (RAN) architectures through software-defined components like xApps and rApps. This paper introduces SAJD, a self-adaptive jammer detection framework that autonomously detects, classifies, and mitigates jamming attacks in O-RAN environments. SAJD forms a closed-loop system by combining real-time inference via xApps with continuous monitoring and retraining through rApps. It uses live telemetry to detect model drift, triggers unsupervised data labeling and model retraining via ClearML, and updates deployed models on the fly, without service disruption. Experiments on a fully O-RAN-compliant testbed show that SAJD outperforms state-of-the-art (offline trained with manual labels) based static ML models in accuracy and adaptability, handling dynamic and previously unseen interference scenarios with resilience.
\end{comment}
\end{abstract}
\vspace{-0.1in}
\section{Introduction}
As 5G networks become central to mission-critical applications, from battlefield communications and public safety to industrial automation, they are increasingly vulnerable to targeted wireless attacks~\cite{Ref:Intr-Pirayesh}. Among the most disruptive of these threats is jamming, where adversaries deliberately inject radio signals into the network’s operating band to degrade or block communication. The widespread availability of software-defined radios (SDRs) has lowered the technical and financial barriers to initiating such attacks. Hence, adversaries can now employ adaptive jamming techniques that adapt to network conditions in real-time.

These attacks are especially dangerous in dynamic environments such as military deployments, where mobility, terrain, and evolving interference patterns make static defenses ineffective. Traditional interference mitigation approaches rely heavily on predefined thresholds or fixed models that assume relatively stable operating conditions. Unfortunately, these approaches often break down when exposed to unseen jamming scenarios or environmental shifts.

This is where the open radio access network (O-RAN) architecture provides a valuable opportunity. By decoupling software from hardware and introducing programmable control layers -- including near-real-time radio access network (RAN) intelligent controller (near-RT RIC) and non-real-time RAN intelligent controller (non-RT RIC) O-RAN enables fine-grained monitoring and closed-loop control of RAN behavior \cite{Ref:Intr-Arnaz} \cite{Ref:Intr-Marinova}. Microservice-based applications such as xApps and rApps can be deployed to intelligently collect \& analyze telemetry, detect anomalies, and trigger mitigation actions \cite{tripathi2025fundamentals}. These capabilities open the door to adaptive, artificial intelligence (AI) / machine learning (ML)-driven interference management directly embedded into the network control plane.

Several research efforts have explored ML-based jamming interference detection in 5G and O-RAN, utilizing RAN key performance indicators (KPIs) such as reference signal received power (RSRP), signal-to-interference-plus-noise ratio (SINR), and block error rate (BLER) to detect anomalies. Jere et al. \cite{Ref:jam-5g} proposed a Bayesian inference-assisted ML technique that exploits cross-layer KPI data collected on a non-standalone (NSA) 5G new radio (NR) testbed to detect coexisting jamming and subtle interference. However, generic ML-based solutions for jamming interference detection considering 5G NR cannot be directly applied to a 5G O-RAN network due to fundamental architectural and operational differences. Besides, O-RAN imposes strict latency constraints for near-RT decisions; hence, ML-based solutions must be re-engineered as xApps or rApps, aligned with O-RAN’s control loop timing, and standardized messaging formats. Recently, Anand et al. \cite{Ref:rw_Anand}  proposed an ML-based xApp that classifies users based on their service priority \& experienced SINR and finally offloads users experiencing severe co-tier interference to nearby femtocell to improve quality of experience (QoE). 
An ML-based rApp is designed in \cite{Ref:rw_Hassan} to monitor long-term trends in environmental telemetry, such as RSRP, SINR, and user distributions, and to manage the lifecycle of the optimization models. Retraining and updates are performed externally through a cloud-native ML framework, and the updated models provide new configuration recommendations to the xApp deployed in the near-RT RIC. This feedback-driven adaptation mechanism enables the system to maintain high inference accuracy and optimized network performance under evolving interference conditions, deployment scenarios, and traffic patterns. In our previous work \cite{Ref:rw_Nathan}, a closed-loop, AI-driven RAN control framework was demonstrated using an O-RAN SDR testbed. A custom interference classification (IC) xApp was developed and deployed within the near-RT RIC, which collected I/Q samples from the RAN over an E2-like interface, generated spectrograms, and used a trained CNN model to detect interference in real-time. Upon detecting interference, the xApp triggered control actions, such as adaptive modulation and coding scheme (MCS) selection, to mitigate interference and enhance network performance.
%Recently, Anand et al. \cite{Ref:rw_Anand} proposed an xApp to reroute users suffering from co-tier interference, whereas Hassan et al. developed a self-organizing xApp/rApp framework for 5G O-RAN campus networks focused on optimizing coverage and capacity. Their approach used an ML-based recommendation system to adjust radio parameters, such as transmit power, in response to real-time network conditions. Deployed on an O-RAN-compliant testbed, the system demonstrated closed-loop behavior and autonomously improved performance.
Current studies hardly consider production-grade adaptive ML systems, and hence, accuracy and effectiveness can quickly degrade in real-world environments where jammer behavior is dynamic and unpredictable. Moreover, these ML-based solutions largely rely on manually annotated data samples and require continuous monitoring and retraining with human intervention in response to changing network conditions or new jamming tactics.
%However, these ML-based solutions largely rely on manually annotated data samples and require continuous monitoring and retraining in response to changing network conditions or new jamming tactics. Current studies hardly consider production-grade adaptive ML systems, and hence, accuracy and effectiveness quickly degrade in real-world environments where jammer behavior is dynamic and unpredictable.

To address this gap, we propose \textit{SAJD — a Self-Adaptive Jammer Detection} framework for 5G O-RAN networks. The SAJD framework is designed to detect and classify %and mitigate 
jamming attacks or jamming interference leveraging an AI/ML-enabled fully closed-loop architecture. It leverages the non-RT RIC for continuous telemetry collection (i.e., network KPIs), data sample annotation for interference detection, and an ML-based interference detector, and the near-RT RIC for real-time inference and control. The SAJD framework, being a seamlessly integrated AI/ML framework (i.e., ClearML framework), includes an ML model management pipeline that automatically trains or retrains the interference detector using newly labeled data, pushes updated models to deployed xApps, and initiates mitigation actions, all without interrupting service. 
%Fig. \ref{fig:sajd} shows the SAJD architecture embedded in the O-RAN ecosystem. Telemetry from RAN are collected and transferred through O1 and E2 interfaces to the non-RT RIC and the near-RT RIC for training and inference, respectively. An ML-powered xApps perform real-time , while rApps define mitigation policies. %These are enforced via the A1 interface, enabling fast, localized responses without breaking O-RAN modularity. 
The main contributions of this work are as follows: 

 $\bullet$ We propose SAJD -- a self-adaptive jammer detection framework tailored for 5G O-RAN networks, which supports automatic data annotation, ML model training, and detection of wireless jamming attacks under dynamic and unseen channel conditions.

$\bullet$ The SAJD framework is implemented as an end-to-end (E2E) O-RAN compliant platform comprising srsRAN, Kafka, InfluxDB, an open-source ClearML framework, and custom RIC components, that are designed and deployed to support closed-loop learning and control workflows.
    
$\bullet$ The proposed SAJD framework %integrates multiple O-RAN microservices to SAJD, including 
includes (i) a \textit{Labeler rApp} based on the Gaussian mixture model (GMM) clustering method for automatic, unsupervised labeling of RAN KPIs; (ii) an \textit{ML-based real-time interference detection xApp}; and (iii) a \textit{Training Manager rApp} that trains an ML model (i.e., ML classifier) as the jamming interference detector, monitors inference accuracy, and orchestrates retraining with the help of the ClearML framework. 
    
$\bullet$ Performance has been evaluated by experimenting on CCI xG Testbed \emph{@}VT \cite{Ref:ccixg} and NextG Lab \emph{@}NC State \cite{Ref:nglab} and discussed elaborately to validate the efficacy of the proposed SAJD framework compared to the state-of-the-art (SOTA) approach, where the static model (i.e., offline trained) based xApp solution is executed in terms of both accuracy and adaptability, enabling real-time response to novel interference patterns without requiring human intervention.

The remainder of this paper is organized as follows. %Section~\ref{sec:related_work} reviews related works, and 
Section~\ref{sec:oran_background} provides background on the O-RAN architecture and its relevant components. 
Section~\ref{sec:sajd_framework} introduces the proposed SAJD framework in detail. Section~\ref{sec:testbed_results} presents our experimental details in the O-RAN testbed and evaluates the performance of the SAJD framework. Finally, Section~\ref{sec:conclusion} concludes the paper and discusses directions for future research.
\vspace{-0.1cm}
\section{O-RAN Background}\label{sec:oran_background}
\vspace{-0.07cm}
The emergence of O-RAN~\cite{tripathi2025fundamentals} represents a fundamental shift in how traditional RANs are designed and deployed. O-RAN aims at solving the challenges associated with traditional RAN by decoupling RAN functions and standardizing open interfaces, enabling a more modular, flexible, and intelligent network architecture. As a result of this disaggregation, the conventional base station is disintegrated into logically separated parts, i.e., the radio unit (RU), distributed unit (DU), and central unit (CU), which can be individually purchased and tailored \cite{Ref:agarwal}. The RU performs radio frequency (RF) processing, analog-to-digital conversion, and interfaces directly with the wireless medium. The DU is responsible for performing real-time physical layer operations such as scheduling and hybrid automatic repeat request (HARQ) processing, while the CU is responsible for higher-layer operations, e.g., radio resource control (RRC), mobility management, and packet data convergence protocol (PDCP). The O-RAN architecture incorporates an intelligent, software-controlled control plane through the RIC framework, where the near-RT RIC enables xApps for real-time RAN optimization, while the non-RT RIC leverages rApps for policy generation, performance tuning, and ML model training based on historical data. Additionally, to achieve interoperability and an open innovation ecosystem as well as to facilitate real-time control, policy exchange, and network management, several open interfaces, i.e., E2 (e.g., between near-RT RIC \& RAN), A1 (e.g., between RICs), and O1 (e.g., between non-RT RIC \& RAN), have been defined by both 3GPP and the O-RAN Alliance.

\section{The SAJD Framework}\label{sec:sajd_framework}
\begin{figure}[t] %[t]
\centerline{\includegraphics[width=0.48\textwidth]{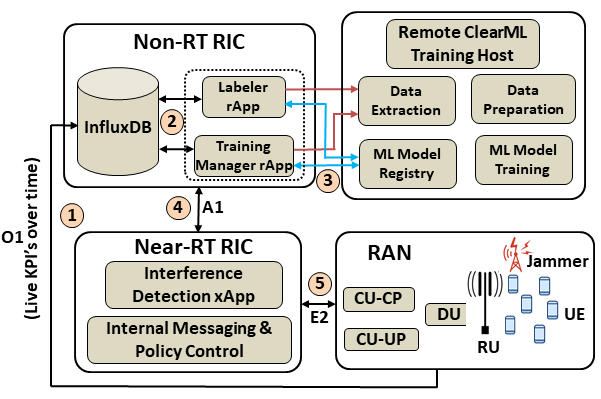}}
\vspace{-0.1in}
\caption{SAJD closed-loop framework.}
\label{fig:sajd}
\vspace{-0.5cm}
\end{figure}

This section introduces the SAJD framework, an automated framework for countering jamming in the O-RAN-based 5G networks. Built on the O-RAN architecture, the SAJD framework leverages ML-driven xApps and rApps to detect interference in real-time. The SAJD framework enables continuous and automated learning from live RAN telemetry using automatically annotated data samples by our developed rApp, executes ML model training/retraining in the ClearML framework, \& deploys updated ML model as xApp—all without human intervention. %This closed-loop system enables it to keep pace with evolving jamming tactics. 
Thus, the SAJD framework provides an E2E scalable and %future-ready 
production-grade AI-powered solution for automatic detection and mitigation in the 5G networks. Fig. \ref{fig:sajd} depicts the SAJD closed-loop O-RAN architecture, and its E2E workflow is discussed below.
\vspace{-0.069cm}
\begin{enumerate}
\item[\textcircled{1}] When the 5G network is operational, live KPIs, including uplink signal-to-noise ratio (UL SNR), uplink modulation and coding scheme (UL MCS), uplink bitrate (UL bitrate), and uplink BLER (UL BLER), are collected from the RAN and sent to InfluxDB, a time-series database deployed in the non-RT RIC, through the O1 interface. These KPIs represent current network conditions and possible jamming interference conditions.  %are reported to the Non-RT RIC over the O1 interface and processed by a consumer module. 
%\item[\textcircled{2}] This module demultiplexes incoming telemetry and writes it to InfluxDB, a time-series database optimized for the speed of ML workloads. 
\item[\textcircled{2}] When a sufficient number of data samples are stored, these data samples are fetched for automatic pre-processing and annotating with the help of the ClearML framework before training/retraining the ML model. The pre-processing and annotating steps are elaborated in the \textit{Labeler rApp} subsection. After annotating the data samples, they are pushed back to the InfluxDB database. A training manager rApp continuously tracks the labeled dataset and monitors ML model performance before initiating the training/retraining pipeline. 

\item[\textcircled{3}] At the beginning, when there is no pretrained ML model, the training manager rApp initiates training by extracting annotated data samples from InfluxDB. After training, the model is stored with its metadata inside the ClearML model registry. Similar steps are also executed if jammer behavior or network behavior drifts as part of continuous monitoring and retraining. Besides, periodic training is also enabled to update the ML model periodically in order to maintain consistent detection performance. 

\item[\textcircled{4}] After completion of the training, when the URL of the latest trained ML model is exposed, the rApp informs the interference detection xApp in the near-RT RIC through the A1 interface of the new model. 
\item[\textcircled{5}] The xApp retrieves the model from the repository and starts real-time inference over live telemetry. Hence, the proposed SAJD closed-loop operation is continuous and autonomous.
\end{enumerate}

In the following subsections, the four major components of the SAJD framework—the labeler rApp, the training manager rApp, the interference detection xApp, and the ClearML training host—are elaborately demonstrated.
\subsection{Labeler rApp} The labeler rApp is designed and developed to annotate raw data samples (e.g., uplink KPIs) irrespective of scenario without the need for human interventions in the closed-loop. Among various KPIs, UL SNR exhibits significant pattern changes when interference is introduced into the 5G network. Therefore, SNR over time is selected as the feature to be annotated in this work. However, the raw SNR values often contain random fluctuations; therefore, it can be challenging to extract meaningful insights accurately. To overcome this, a moving average is applied to smooth out unwanted fluctuations and noise \cite{Ref:amz}. At first, the convolution operation is applied using a uniform kernel of window size $W$, where each value is weighted equally by $\frac{1}{W}$. This operation produces a smoothed signal by averaging values over a sliding window. To maintain the original dataset length, symmetric padding is applied at both ends using edge values. Next, batches of 30 smoothed raw data samples are retrieved successively by the labeler rApp and processed using the standard scaling method. This method normalizes the raw SNR values to have a mean of $0$ and a standard deviation of $1$, ensuring consistent scaling of SNR data samples across various scenarios. For each batch, the average rate of change ($ARC$) is calculated as follows:
\begin{equation}
\textit{ARC} = \frac{1}{N-1} \sum_{i=1}^{N-1} \left( \text{data}[i+1] - \text{data}[i] \right)
\end{equation}
The $ARC$ is calculated to ensure that data samples of two classes are prepared for training the GMM clustering method for a given network scenario.  An empirically chosen threshold,~$\tau = 0.0004$, is defined to trigger GMM training. Unlike other hard clustering methods, such as K-Means, GMM performs soft clustering, considering a cluster of data as a Gaussian distribution with its mean and covariance. Utilizing statistical features, such as covariance and mean, the GMM can effectively capture the inner relationship and each data point to a cluster based on the higher probability \cite{Ref:yli}. However, when the training of the GMM model is triggered, the GMM model starts to generalize to unseen data and converges to perform clustering based on probability. The entire workflow of the labeler rApp at time step $t$ is presented in Algorithm \ref{alg:adaptive_gmm}.     
\begin{algorithm}[t]
\small
\caption{Automatic annotating of raw data samples using GMM clustering technique.}
\label{alg:adaptive_gmm}
\begin{algorithmic}[1]
 \renewcommand{\algorithmicrequire}{\textbf{Input:}}
 \renewcommand{\algorithmicensure}{\textbf{Output:}}
\REQUIRE A batch of smoothed SNR Data samples $\mathcal{D}_t = \{x_1, x_2, \dots, x_N\}$, a threshold $\tau$; 

\ENSURE Cluster labels $\mathcal{L}_t = \{0, 1\}$.

\STATE \textbf{Initialize:} A batch of smoothed SNR data samples $\mathcal{D}_t$, $start\_idx \gets 0$, $last\_change\_idx \gets 0$, Cluster labels $\mathcal{L}_t = \{ \}$.
\STATE $end\_idx \gets start\_idx + length\_of\_batch \_size$.
\STATE Compute average rate of change (ARC).

\IF{$|arc| > \tau$}
    \STATE Process the data for training.
    \STATE Train a new GMM model $\text{GMM}_t$ on the processed data:
    \[
    \text{GMM}_t \gets \text{Train}(\mathcal{D}_t)
    \]
    \STATE Predict cluster labels $\mathcal{L}_t$ using $\text{GMM}_t$:
    \[
    \mathcal{L}_t \gets \text{Predict}(\text{GMM}_t, \mathcal{D}_t)
    \]
    \STATE Update $last\_change\_idx \gets end\_idx$
\ENDIF

\begin{comment}
    \STATE Check if a GMM model is already trained.

\STATE \textbf{Spike Detection:} Analyze $\mathcal{D}_t$ for detecting thin spikes.

\STATE Define a spike as a region with unusually high density over a narrow range.

\STATE \textbf{Decision Process:}
\IF {spikes are detected in $\mathcal{D}_t$ \textbf{or} a trained GMM model is not found}
    \STATE Process the data for training.
    \STATE Train a new GMM model $\text{GMM}_t$ on the processed data:
    \[
    \text{GMM}_t \gets \text{Train}(\mathcal{D}_t)
    \]
    \STATE Predict cluster labels $\mathcal{L}_t$ using $\text{GMM}_t$:
    \[
    \mathcal{L}_t \gets \text{Predict}(\text{GMM}_t, \mathcal{D}_t)
    \]
    \STATE Update $last\_change\_idx \gets end\_idx$
\ELSE
    \STATE Use the previously trained GMM model $\text{GMM}_{t-1}$ to predict labels:
    \[
    \mathcal{L}_t \gets \text{Predict}(\text{GMM}_{t-1}, \mathcal{D}_t)
    \]
\ENDIF
\end{comment}

\STATE \textbf{Output:} Return cluster labels $\mathcal{L}_t$.
\end{algorithmic}
\end{algorithm}

\subsection{Training Manager rApp} The Training Manager rApp is another key component of our closed-loop SAJD framework, providing full-lineage traceability over the lifecycle of the ML model, incorporating real-time network feedback. Its primary responsibility is to initiate training by communicating with the ClearML ML model training module when there is no trained ML model and to continuously monitor ML model prediction performance. If ML model detection performance falls behind $30\%$,  it dictates automatic retraining with the most recent annotated KPIs. When the training/retraining is required, it clones a template task describing the training procedure, submits it to a ClearML queue, monitors it, and once completed, it notifies the interference detection xApp using an A1-like interface with the task ID of the newly trained model for continuous deployment. Besides, it manages task orchestration, remote execution, and experiment tracking. In addition, this rApp also enables periodic training of the ML model to ensure consistent performance in case of slight data drifting in the same scenario. %\hl{A lightweight $3$ layers dense neural network-based ML model is used to detect jamming interference. The model layers consisting of $32$, $16$, and $8$ neurons along with the rectified linear unit (ReLU) activated as well as extract and encode features from the $60$-dimensional input vector derived from a sliding window spanning $15$ consecutive time steps, with each step comprising four KPI values described in Section} \ref{sec:sajd_framework}. 
A lightweight three-layer dense neural network is employed to detect jamming interference. The layers contain 32, 16, and 8 neurons, respectively, each activated with the rectified linear unit (ReLU) function to extract and encode features. The model takes as input a 60-dimensional vector derived from a sliding window of 15 consecutive time steps, where each step comprises four KPI values, as described in Section \ref{sec:sajd_framework}.
The output layer comprises two neurons, each corresponding to the classes "no interference" and "interference", and uses a softmax activation function to produce class probabilities. The ML model is trained using the "Root Mean Square Propagation (RMSprop)" optimizer with a learning rate of $0.01$ for $50$ iterations, utilizing a mini-batch size of $64$.

\subsection{Interference Detection xApp} The interference detection xApp is a closed-loop real-time control application designed to detect and remediate wireless interference in O-RAN networks using ML-based control decisions. Its design combines E2 telemetry processing and A1 model management interfaces to achieve an E2E coupling between sensing, inference, and RAN reconfiguration. The xApp retrieves KPIs from the uplink channel, such as SNR, bitrate, BLER, and MCS. First, the xApp receives streaming KPI data over an SCTP-based E2-like interface, which is used as input to a dense neural network model developed for binary detection of the presence of interference. The xApp logs the model's inferences, which indicate whether interference was present on the uplink channel at a given time.
The model can be updated in-place via an HTTP-based A1-like interface, from which a non-RT rApp in ClearML can inform the xApp that the new model is trained and ready. This architecture enables the xApp to perform any-time updates of models without disrupting service while adapting over time to changes in interfering patterns. The inference outputs readily dictate radio resource control, e.g., the possibility of dynamically selecting MCS to implement autonomous interference mitigation policies. This architecture is motivated by preceding research \cite{Ref:rw_Nathan} around intelligent xApps and RIC-driven anomaly detection and represents the realization of ML-reinforced RRM in O-RAN networks.
\vspace{-2mm}
\subsection{ClearML Training Host}The ClearML framework is an E2E machine learning operations (MLOps) engine for deploying and monitoring ML models \cite{Ref:clearml}. It comprises three essential layers: the infrastructure control plane (ICP), the AI development center, and the GenAI App engine. %In the ClearML framework, every experiment is considered a task.
A ClearML agent (i.e., one of the sub-components of ICP) is launched to prepare the ML infrastructure, which acts as a virtual environment and execution manager for an ordered list of scheduled tasks being integrated with the ClearML server. In our work, the ClearML framework is used to leverage a production-grade ML pipeline that includes training data collection, training data preprocessing, ML model training/retraining, and model versioning \& storing.

\section{EXPERIMENTATION AND PERFORMANCE ANALYSIS}\label{sec:testbed_results}
To evaluate the efficacy of the proposed SAJD framework, we have created an O-RAN-based network using the srsRAN radio protocol stack~\cite{Ref:srsran} integrated with the core network on an O-RAN-compliant testbed at the NextG Lab @NC State.
%In order to evaluate the efficacy of the proposed SAJD framework, we have created an O-RAN-based network using the srsRAN radio protocol stack \cite{Ref:srsran} integrated with core network. 
Here, the xApp and rApps are designed to be deployable as containers in the O-RAN software community near-RT RIC and non-RT RICs. In addition, the O1 interface is managed by a Kafka broker where a modified version of srsRAN acts as a Kafka producer, and a service in the non-RT RIC acts as a Kafka consumer. The ClearML training host is accessible to the services in the near-RT and non-RT RICs. When the network is operational, the transmit signal power is set to $-4.5$ dB, a $20$ Mbps data rate is observed, and the MCS is set to 24. We tested the system under varying channel conditions and interference levels when the data was collected. Table \ref{tab:interference_noise} depicts the parameters considered in this study to introduce interference and non-interference scenarios (Scen.).

\begin{comment}
\begin{table}[t]
\caption{Interference (Int.) and Noise Parameters}
\label{tab:interference_noise}
\centering
\small % Reduce font size
\setlength{\tabcolsep}{4pt} % Reduce column spacing

\begin{tabular}{@{}cccc@{}}
\toprule
\textbf{Scenario} & \textbf{Int. Event} & \textbf{Int. (dB)} & \textbf{Noise Amplitude} \\ \midrule
1  & ON  & -8   & 0.056 \\
2  & OFF & -100 & 0.056 \\
3  & ON  & -8   & 0.15  \\
4  & OFF & -100 & 0.15  \\
5  & ON  & -8   & 0.33  \\
6  & OFF & -100 & 0.33  \\
7  & ON  & -20  & 0.056 \\
8  & OFF & -100 & 0.056 \\
9  & ON  & -20  & 0.15  \\
10 & OFF & -100 & 0.15  \\
11 & ON  & -20  & 0.33  \\
12 & OFF & -100 & 0.33  \\
13 & ON  & -40  & 0.056 \\
14 & OFF & -100 & 0.056 \\
15 & ON  & -40  & 0.15  \\
16 & OFF & -100 & 0.15  \\
17 & ON  & -40  & 0.33  \\
18 & OFF & -100 & 0.33  \\ \bottomrule
\end{tabular}
\vspace{-4mm}
\end{table}
\end{comment}

\begin{table}[h]

\footnotesize
\caption{Interference (Int.) and noise parameters.}
\label{tab:interference_noise}
%\small % Reduce font size
%\setlength{\tabcolsep}{-2pt} % Reduce column spacing    
\centering
\begin{tabular}{@{}c c c c c c c c@{}}
\toprule
\textbf{Scen.} & \makecell{\textbf{Int.} \\ \textbf{Event}} & \makecell{\textbf{Int.} \\ \textbf{(dB)}} & \makecell{\textbf{Noise} \\ \textbf{Amp.}} &
\textbf{Scen.} & \makecell{\textbf{Int.} \\ \textbf{Event}} & \makecell{\textbf{Int.} \\ \textbf{(dB)}} & \makecell{\textbf{Noise} \\ \textbf{Amp.}} \\
\midrule
1  & ON   & -8   & 0.056 & 10 & OFF & -100 & 0.15  \\
2  & OFF  & -100 & 0.056 & 11 & ON  & -20  & 0.33  \\
3  & ON   & -8   & 0.15  & 12 & OFF & -100 & 0.33  \\
4  & OFF  & -100 & 0.15  & 13 & ON  & -40  & 0.056 \\
5  & ON   & -8   & 0.33  & 14 & OFF & -100 & 0.056 \\
6  & OFF  & -100 & 0.33  & 15 & ON  & -40  & 0.15  \\
7  & ON   & -20  & 0.056 & 16 & OFF & -100 & 0.15  \\
8  & OFF  & -100 & 0.056 & 17 & ON  & -40  & 0.33  \\
9  & ON   & -20  & 0.15  & 18 & OFF & -100 & 0.33  \\
\bottomrule
\end{tabular}
\vspace{-1mm}
\end{table}

\begin{comment}
\begin{table}[h]
\centering
\begin{tabular}{@{}c@{\hskip 6pt}c@{\hskip 6pt}c@{\hskip 6pt}c@{\hskip 12pt}c@{\hskip 6pt}c@{\hskip 6pt}c@{\hskip 6pt}c@{}} 
\toprule
\textbf{Scenario} & \shortstack{\textbf{Int.} \\ \textbf{Event}} & \shortstack{\textbf{Int.} \\ \textbf{(dB)}} & \shortstack{\textbf{Noise} \\ \textbf{Amp.}} & 
\textbf{Scenario} & \shortstack{\textbf{Int.} \\ \textbf{Event}} & \shortstack{\textbf{Int.} \\ \textbf{(dB)}} & \shortstack{\textbf{Noise} \\ \textbf{Amp.}} \\ 
\midrule
1  & ON   & -8   & 0.056 & 10 & OFF & -100 & 0.15  \\
2  & OFF  & -100 & 0.056 & 11 & ON  & -20  & 0.33  \\
3  & ON   & -8   & 0.15  & 12 & OFF & -100 & 0.33  \\
4  & OFF  & -100 & 0.15  & 13 & ON  & -40  & 0.056 \\
5  & ON   & -8   & 0.33  & 14 & OFF & -100 & 0.056 \\
6  & OFF  & -100 & 0.33  & 15 & ON  & -40  & 0.15  \\
7  & ON   & -20  & 0.056 & 16 & OFF & -100 & 0.15  \\
8  & OFF  & -100 & 0.056 & 17 & ON  & -40  & 0.33  \\
9  & ON   & -20  & 0.15  & 18 & OFF & -100 & 0.33  \\
\bottomrule
\end{tabular}
\vspace{-2mm}
\caption{Scenarios showing interference status, signal degradation, and noise amplitude.}
\end{table}
\end{comment}

\begin{figure}[t]
  \centering
  \begin{subfigure}[b]{0.43\textwidth}
    \includegraphics[width=\textwidth]{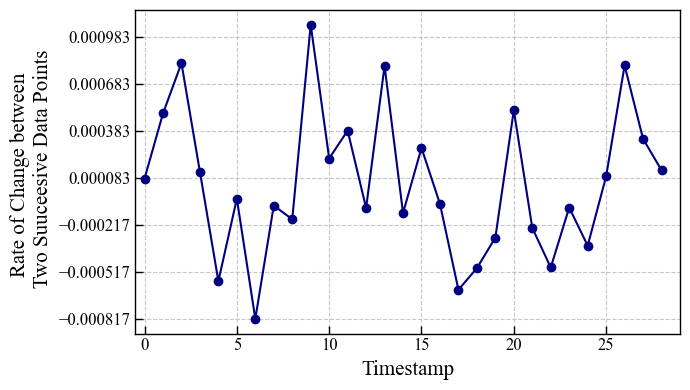}
   \caption{When the data points are from same events.}
    \label{fig:rc1}
  \end{subfigure}
  %\hfill
  \begin{subfigure}[b]{0.40\textwidth}
    \includegraphics[width=\textwidth]{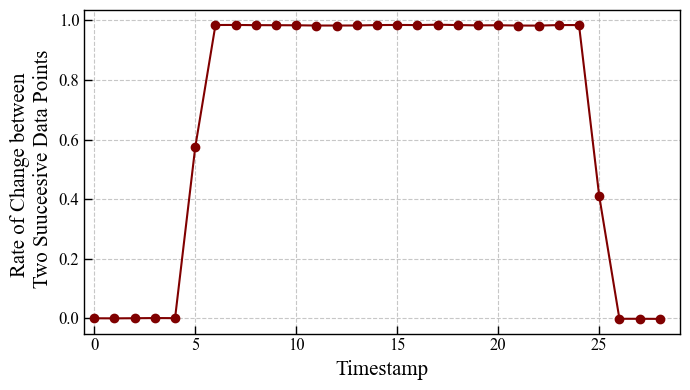}
    \caption{When the data points are from two different events.}
    \label{rc-2}
  \end{subfigure}
  \caption{Illustration of the rate of change between two successive data points.}
  \label{fig:rccomp}
\end{figure}

\begin{figure}[t]
    \centering
\includegraphics[width=.45\textwidth]{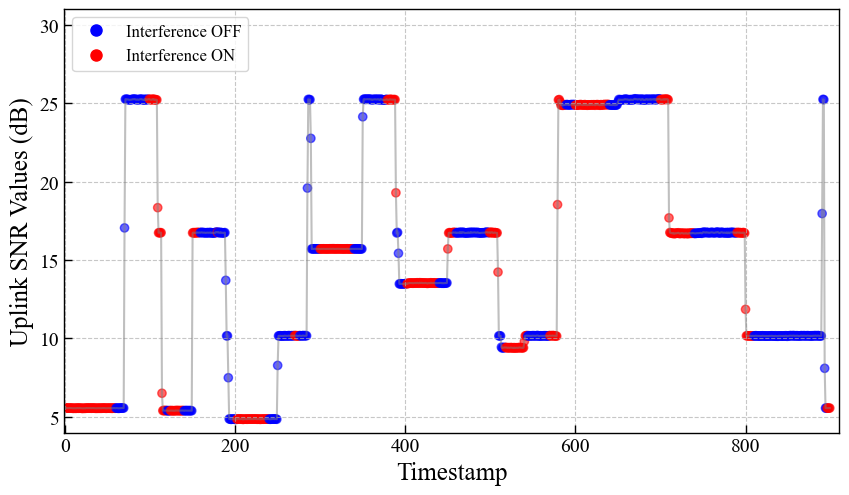}
    \caption{Labeler rApp prediction performance for different scenarios.}
    \label{rapp-res}
    \vspace{-0.7cm}
\end{figure}

Before deploying the labeler rApp in the real-time closed-loop system at the NextG Lab Testbed, its offline efficacy was validated using collected KPIs across various scenarios listed in Table \ref{tab:interference_noise} at the CCI xG Testbed. Fig. \ref{fig:rccomp} compares the rate of change between successive data points for one same-event and one different-event batch, respectively. After analyzing all the batches formed using collected data based on Table \ref{tab:interference_noise}, the threshold value $ARC$ has been determined. Fig. \ref{rapp-res} demonstrates satisfactory performance, with minor mislabeling near event transitions. 

\begin{figure}[t]
  \centering
  \begin{subfigure}[b]{0.48\textwidth}
    \includegraphics[width=\textwidth, trim=0cm 0.4cm 0cm 0.35cm, clip]{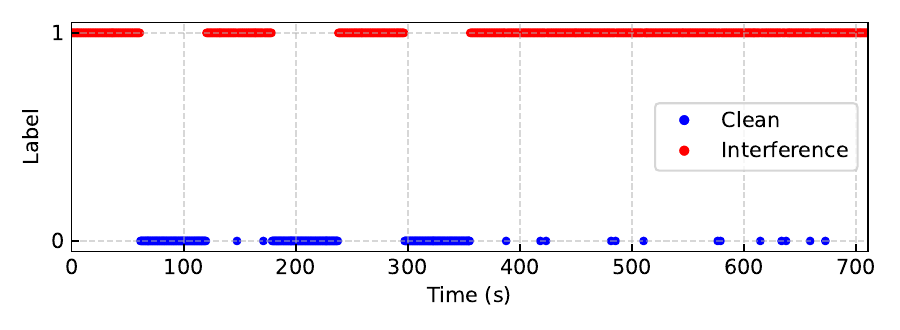}
    \caption{SOTA approach (offline-trained).}
    \label{fig:baseline}
  \end{subfigure}
  %\hfill
  \begin{subfigure}[b]{0.48\textwidth}
    \includegraphics[width=\textwidth, trim=0cm 0.4cm 0cm 0cm, clip]{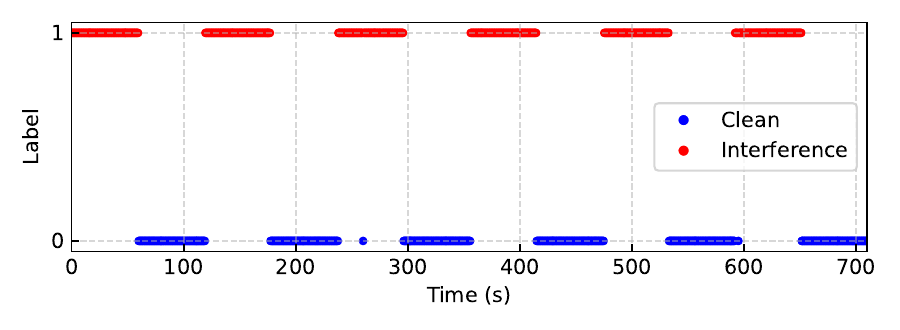}
    \caption{SAJD approach (online-retrained).}
    \label{fig:retrained}
  \end{subfigure}
  \caption{Comparison of interference detection between the SAJD and the SOTA approach under two different scenarios.}
  \label{fig:xapp_ic_comparison}
  \vspace{-4mm}
\end{figure}

\begin{figure}[htbp] %[t]
\centerline{\includegraphics[width=0.48\textwidth]{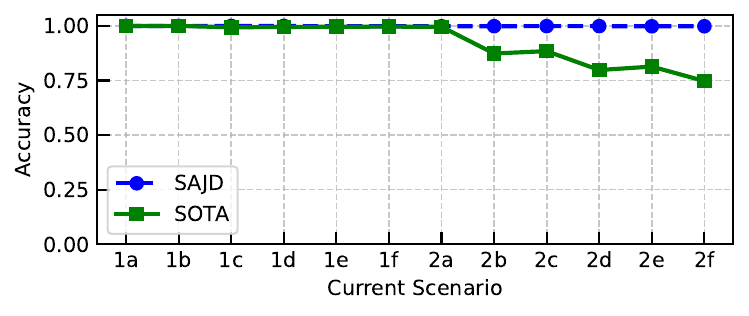}}
%\includegraphics[width=\textwidth, trim=0cm 0.4cm 0cm 0.35cm, clip]{sota_ic_prediction_fixed.pdf}
%\caption{Accuracy over time for SOTA and SAJD models during dynamic interference conditions.}
%\caption{Temporal classification accuracy of interference detection using SOTA and SAJD models.}
\caption{Comparison of interference detection accuracy across windows of sequential scenes (1a–2f).}
%Scenario 1 (1a–1f) is the conditions viewed during training, and Scenario 2 (2a–2f) is the unseen scenes. The SAJD model has stable performance for both scenes, while the SOTA model has a noticeable decline in unseen scenes.
\label{fig:accuracy}
\vspace{-0.2cm}
\end{figure}

After validation, % the labeler rApp on offline UL SNR data using the CCI xG Testbed, 
the labeler rApp was deployed in the live NextG Lab Testbed to support real-time closed-loop training, retraining, and interference detection. %After validation on offline data in the CCI xG Testbed, the labeler was deployed in the non-RT RIC during live network operation in the NextG Lab Testbed to annotate raw data samples from the InfluxDB database}. 
~Initially, the training manager rApp initiated fresh ML model training by retrieving a large amount of annotated data through the ClearML framework and then notified the interference detection xApp via the A1-like interface with the URL of the trained model.
%Fig. \ref{fig:rccomp} presents a comparison of the rate of change variation over time between two successive data samples for two batches: one batch includes data points from the same events, and the other one includes data points from two different events. After analyzing all the batches formed using collected data based on Table \ref{tab:interference_noise}, the threshold value $ARC$ has been determined. Next, the prediction performance of the labeler rApp is presented in Fig. \ref{rapp-res}. The labeler rApp seems to annotate well, except for some incorrectness at the edges of transitions.     
%After validating the labeler rApp performance on offline data, it was deployed in the non-RT RIC. When the network begins to operate, the raw data samples are collected and stored in the InfluxDB database. Since there were no pretrained ML classifiers at the beginning, the training manager replaced ML classifier training by retrieving a sufficient amount of annotated data from the labeler rApp. When the ML classifier training is finished, ClearML exposes a URL to download the newly trained ML model, and the training manager rApp notifies the interference detection xApp using the A1-like interface, forwarding the model's URL. After receiving the trained ML classifier, the xApp starts detecting interference presence in the network.
%\vspace{-2mm}
\begin{comment}
  \begin{table}[h]
\caption{Interference (Int.) and noise parameters setup during ML model performance evaluation.}
\label{tab:interference_noise_2}
%\small % Reduce font size
%\setlength{\tabcolsep}{-2pt} % Reduce column spacing    
\centering
\begin{tabular}{@{}c c c c c c c c@{}}
\toprule
\textbf{Scen.} & \makecell{\textbf{Int.} \\ \textbf{Event}} & \makecell{\textbf{Int.} \\ \textbf{(dB)}} & \makecell{\textbf{Noise} \\ \textbf{Amp.}} &
\textbf{Scen.} & \makecell{\textbf{Int.} \\ \textbf{Event}} & \makecell{\textbf{Int.} \\ \textbf{(dB)}} & \makecell{\textbf{Noise} \\ \textbf{Amp.}} \\
\midrule
1a  & ON   & -12   & 0.1 & 2a & ON & -12 & 0.333  \\
1b  & OFF  & -100 & 0.1 & 2b & OFF & -100  & 0.333  \\
1c  & ON   & -12   & 0.1  & 2c & ON & -12 & 0.333  \\
1d  & OFF  & -100 & 0.1  & 2d & OFF  & -100  & 0.333 \\
1e  & ON   & -12   & 0.1  & 2e & ON & -12 & 0.333 \\
1f  & OFF  & -100 & 0.1  & 2f & OFF & -100  & 0.333  \\
\bottomrule
\end{tabular}
\vspace{-2mm}
\end{table}    
\end{comment}
Next, the xApp started to detect jamming interference utilizing the trained model. When there's a change in the scenario, the SAJD framework autonomously detected performance degradation, prompting the rApp to initiate retraining using newly labeled telemetry data and release the updated model to the xApp without causing any service interruption. The performance of the SAJD framework is evaluated at NextG Lab Testbed on the $12$ scenarios, where the first six scenarios were set with a noise amplitude of 0.1 and interference power level varied between $–12$~dB and $–100$~dB, while the next six scenarios were configured with a higher noise amplitude of $0.333$ under the same interference conditions. Fig. \ref{fig:xapp_ic_comparison} depicts the interference detection performance of the proposed SAJD framework compared to the SOTA approach. %(i.e., offline-trained interference detection xApp) under two different scenarios with noise amplitude $0.1$ and $0.333$. 
In Fig. \ref{fig:baseline}, when the networking scenario changes between the $350$th to $400$th seconds, the interference detection performance becomes inconsistent and hence starts decreasing. In contrast, Fig.~\ref{fig:retrained} demonstrates that the SAJD framework adapts to new scenarios, and the interference detection performance remains consistent in real-time without interruption. Since the SAJD framework enables a continuous monitoring and retraining pipeline with the help of automatically annotated data, it ensures robust and fast adaptive behavior without any human interventions in the loop. Fig. \ref{fig:accuracy} compares the accuracy of the SAJD and SOTA approaches across sequential scenario windows in an O-RAN network under dynamically changing interference. The SAJD approach achieves consistently high accuracy during the 12-minute test session, even for interference transitions and unseen situations. This satisfies its strong generalization capability and robustness, suitable for deployment as a real-time xApp in the near-RT RIC. The baseline SOTA approach, however, shows considerable accuracy drops, especially when confronted with unknown interference patterns. This indicates weak robustness of the SOTA approach in unseen environments (2a–2f), whereas the SAJD approach demonstrates its capability to detect interference events in real-time O-RAN networks.
\vspace{-2mm}

%\vspace{-2mm}
\section{Conclusion}\label{sec:conclusion}
%\vspace{-1mm}
In this paper, we introduce a self-adaptive jammer detection (SAJD) framework for 5G O-RAN systems that combines ML-driven rApps and xApps as well as incorporates a closed-loop pipeline that supports automatic labeling, training/retraining ML model, and detecting jamming interference in the dynamic network scenarios. Through periodic KPI gathering, continuous model retraining, and transparent model deployment, the SAJD framework enhances the accuracy of the interference detection in unseen channel conditions. The findings verify that the SAJD framework is not only intelligently adaptive but also scalable, efficient, and a future-proof interference management solution for the next-generation RAN operations. Nonetheless, the SAJD framework can be vulnerable to data poisoning or other adversarial attacks; therefore, further study on susceptibility to adversarial attacks will be conducted in the future.% as an extension of this work.

%While SAJD is resilient to dynamic jamming, its automatic labeling and retraining pipeline can potentially be compromised by adversarial manipulation or data poisoning. In our future work, we will investigate defenses including adversarial training, minimal human-in-the-loop verification, and secure provenance tracking to enhance resilience in high-security environments. Moreover, we plan to classify multiple interference classes and the autonomous action of interference mitigation.

\section*{Acknowledgment}
%This work was funded by Commonwealth Cyber Initiative (CCI)/ CCI xG Testbed. Visit CCI at: \url{www.cyberinitiative.org} and \url{www.ccixgtestbed.org}.
This work has been supported by the NSF awards \#2120411 and \#2443035, and by the Commonwealth Cyber Initiative (CCI)/ CCI xG Testbed. Visit CCI at: \url{www.cyberinitiative.org} and \url{www.ccixgtestbed.org}.

\bibliographystyle{ieeetr}
%\bibliography{ref}
\bibliography{main}

\begin{thebibliography}{10}

\bibitem{Ref:Intr-Pirayesh}
H.~Pirayesh and H.~Zeng, ``Jamming attacks and anti-jamming strategies in wireless networks: A comprehensive survey,'' {\em IEEE Comm. Sur. \& Tut.}, vol.~24, no.~2, pp.~767--809, 2022.

\bibitem{Ref:Intr-Arnaz}
A.~Arnaz, J.~Lipman, M.~Abolhasan, and M.~Hiltunen, ``Toward integrating intelligence and programmability in open radio access networks: A comprehensive survey,'' {\em IEEE Access}, vol.~10, pp.~67747--67770, 2022.

\bibitem{Ref:Intr-Marinova}
S.~Marinova and A.~Leon-Garcia, ``Intelligent {O-RAN} beyond {5G}: Architecture, use cases, challenges, and opportunities,'' {\em IEEE Access}, vol.~12, pp.~27088--27114, 2024.

\bibitem{tripathi2025fundamentals}
N.~D. Tripathi and V.~K. Shah, {\em Fundamentals of {O-RAN}}.
\newblock Hoboken, NJ, USA: Wiley-IEEE Press, Feb. 2025.

\bibitem{Ref:jam-5g}
S.~Jere, Y.~Wang, I.~Aryendu, S.~Dayekh, and L.~Liu, ``Bayesian inference-assisted machine learning for near real-time jamming detection and classification in 5{G} new radio,'' {\em IEEE Trans. on Wireless Comm.}

\bibitem{Ref:rw_Anand}
D.~Anand, M.~A. Togou, and G.~M. Muntean, ``A machine learning-based {xAPP} for {5G O-RAN} to mitigate co-tier interference and improve {QoE} for various services in a {HetNet} environment,'' in {\em IEEE Int. Sym. on Broadband Multimedia Sys. and Broadcasting (BMSB)}, pp.~1--6, 2023.

\bibitem{Ref:rw_Hassan}
M.~Hassan, A.~Diab, S.~Parameswaran, and A.~Mitschele-Thiel, ``Machine learning-based coverage and capacity optimization {xApp/rApp} for open {RAN 5G} campus networks,'' in {\em Mobilkommunikation; 28. ITG-Fachtagung}, (Osnabrück), pp.~191--196, 2024.

\bibitem{Ref:rw_Nathan}
N.~H. Stephenson, A.~J. Chiejina, N.~B. Kabigting, and V.~K. Shah, ``Demonstration of closed loop {AI}-driven {RAN} controllers using {O-RAN SDR} testbed,'' in {\em IEEE Military Comm. Conf. (MILCOM)}, (Boston, MA, USA), pp.~241--242, 2023.

\bibitem{Ref:ccixg}
A.~D. Silva, M.~R. Chowdhury, A.~Sathish, A.~Tripathi, S.~F. Midkiff, and L.~A.~D. Silva, ``{CCI xG} testbed: An {O-RAN} based platform for future wireless network experimentation,'' {\em IEEE Comm. Mag.}, vol.~63, no.~2, pp.~62--68, 2025.

\bibitem{Ref:nglab}
``{NextG Lab @ NC State}.'' \url{https://www.nextgwirelesslab.org}, 2021.
\newblock Accessed August 10, 2025.

\bibitem{Ref:agarwal}
B.~Agarwal, R.~Irmer, D.~Lister, and G.-M. Muntean, ``Open {RAN} for {6G} networks: Architecture, use cases and open issues,'' {\em IEEE Comm. Sur. \& Tut.}, 2025.

\bibitem{Ref:amz}
A.~M. Zulhakim, W.~F.~H. Abdullah, I.~S.~A. Halim, R.~B.~H. Mamat, M.~I.~A. Muslan, and A.~Z.~A. Bakar, ``Smoothing sensor data in a controlled {IoT} framework with moving averages,'' in {\em Proc. 2023 IEEE Regional Symp. on Micro and Nanoelectronics (RSM)}, (Langkawi, Malaysia), pp.~86--89, 2023.

\bibitem{Ref:yli}
Y.~Li, J.~Zhang, Z.~Ma, and Y.~Zhang, ``Clustering analysis in the wireless propagation channel with a variational gaussian mixture model,'' {\em IEEE Trans. on Big Data}, vol.~6, no.~2, pp.~223--232, 2020.

\bibitem{Ref:clearml}
``What is {ClearML?}.'' \url{https://clear.ml/docs/latest/docs/}, 2025.
\newblock Accessed August 10, 2025.

\bibitem{Ref:srsran}
``{Open source 5G software suite srsRAN project}.'' \url{https://github.com/srsran/srsRAN}.
\newblock Accessed August 10, 2025.

\end{thebibliography}

\end{document}